\documentclass[conference, hidelinks]{IEEEtran}
\IEEEoverridecommandlockouts

\RequirePackage{cite}
\RequirePackage{amsmath,amssymb,amsfonts}
\RequirePackage{algorithmic}
\RequirePackage{graphicx}
\RequirePackage{float}
\RequirePackage[caption=false]{subfig}
\RequirePackage{textcomp}
\RequirePackage[dvipsnames]{xcolor}
\RequirePackage{orcidlink}
\RequirePackage[acronym]{glossaries}
\RequirePackage[detect-all=true,per-mode=symbol,per-symbol =/]{siunitx}
\RequirePackage[capitalise]{cleveref}
\RequirePackage{tablefootnote}
\RequirePackage{threeparttable}
\RequirePackage{booktabs}
\RequirePackage{listings}
\RequirePackage{graphicx}

\DeclareSIUnit{\x}{\!\ensuremath{\times}}
\DeclareSIUnit\gateeq{GE}
\newcommand*\circled[1]{\tikz[baseline=(char.base)]{
		\node[shape=circle,draw,inner sep=1pt] (char) {#1};}}

\ifdefined\blind
    \newcommand{\basilisk}{\textcolor{red}{\scalebox{.6}[1.0]{\textit{BLINDNAME}}}}
    \newcommand{\pulpwarecite}{\footnote{Citation omitted for blind review.}}
    \newcommand{\basiliskcite}{\footnote{Citation omitted for blind review.}}
    \newcommand{\yosysthanksnote}{\footnote{Acknowledgment omitted for blind review.}}
\else
    \newcommand{\basilisk}{Basilisk}
    \newcommand{\pulpwarecite}{\cite{pulpware-git}}
    \newcommand{\basiliskcite}{\cite{basilisk-git}}
    \newcommand{\abcthanksnote}{\footnote{The help and advice of Alan Mishchenko, Masahiro Fujita, Giovanni De Micheli, Andrea Costamagna, Alessandro Tempia Calvino is gratefully acknowledged.}}
    \newcommand{\yosysthanksnote}{\footnote{The authors highly acknowledge the support by the Yosys team and especially Martin Povišer with the Yosys implementation.}}
\fi

\lstdefinelanguage{other}{
    basicstyle=\itshape,
}

\lstdefinelanguage{abc}{
    language=tcl,
    columns=fullflexible,
    keepspaces=true,
    upquote=true,
    showstringspaces=false,
    basicstyle=\normalsize,
    commentstyle=\color{gray},
    keywordstyle=\color{Blue}\bfseries,
    morekeywords={&get, &st, &if, &syn2, &b, &dch, &put, &nf, rec_add3, rec_start3, read_lib}
}

\lstdefinelanguage{yosys}{
    language=tcl,
    columns=fullflexible,
    keepspaces=true,
    upquote=true,
    showstringspaces=false,
    basicstyle=\normalsize,
    commentstyle=\color{gray},
    keywordstyle=\color{Sepia}\bfseries,
    morekeywords={booth, maccmap, alumacc, extract, abc, abc9}
}

\newacronym{qor}{QoR}{quality of results}
\newacronym{soc}{SoC}{system-on-chip}
\newacronym{rtl}{RTL}{register transfer level}
\newacronym{pnr}{P\&R}{place and route}
\newacronym{eda}{EDA}{electronic design automation}
\newacronym{ast}{AST}{abstract syntax tree}
\newacronym{rtlil}{RTLIL}{RTL intermediate language}
\newacronym{asic}{ASIC}{application-specific integrated circuit}
\newacronym{macc}{MACC}{multiply-accumulate}
\newacronym{fma}{FMA}{fused multiply-add}
\newacronym{drc}{DRC}{design rule check}
\newacronym{nda}{NDA}{non-disclosure agreement}
\newacronym{mpw}{MPW}{multi-project wafer}
\newacronym{csa}{CSA}{carry-save adder}
\newacronym{cpa}{CPA}{carry-propagate adder}
\newacronym{cia}{CIA}{carry-increment adder}
\newacronym{rca}{RCA}{ripple-carry adder}
\newacronym{ppa-bk}{PPA-BK}{Brent-Kung parallel-prefix adder}
\newacronym{ppa-sk}{PPA-SK}{Sklansky parallel-prefix adder}
\newacronym{cmos}{CMOS}{complementary metal-oxide semiconductor}
\newacronym{aig}{AIG}{and-inverter graph}
\newacronym{lms}{LMS}{lazy man's synthesis}
\newacronym{orfs}{ORFS}{OpenROAD flow scripts}
\newacronym{pdk}{PDK}{process design kit}
\newacronym{fsm}{FSM}{finite-state machine}
\newacronym{fpu}{FPU}{floating-point unit}
\newacronym{lec}{LEC}{logic equivalence checking}
\newacronym{hitl}{HITL}{human-in-the-loop}
\newacronym{fpga}{FPGA}{field-programmable gate array}
\newacronym{lut}{LUT}{lookup table}
\newacronym{oseda}{OS EDA}{open-source EDA}

\newcommand\riscv{\mbox{RISC-V}}
\newcommand\sv{\mbox{SystemVerilog}}
\newcommand\rtg{\gls{rtl}-to-\mbox{GDSII}}
\newcommand\rtgn{\gls{rtl} to \mbox{GDSII}}

\title{Insights from {\basilisk}: Are Open-Source EDA Tools Ready for a Multi-Million-Gate, Linux-Booting RV64 SoC Design?}

\begin{document}

\ifdefined\blind
    \author{%
    \vspace{0.1cm} %
    \textit{Authors omitted for blind review}
    \vspace{0.1cm} %
    }
\else
\author{
    \IEEEauthorblockN{%
    Philippe Sauter\orcidlink{0009-0001-6504-8086}\IEEEauthorrefmark{1}\IEEEauthorrefmark{10}, %
    Thomas Benz\orcidlink{0000-0002-0326-9676}\IEEEauthorrefmark{1}\IEEEauthorrefmark{10}, %
    Paul Scheffler\orcidlink{0000-0003-4230-1381}\IEEEauthorrefmark{1}, %
    Frank K. Gürkaynak\orcidlink{0000-0002-8476-554X}\IEEEauthorrefmark{1}, %
    Luca Benini\orcidlink{0000-0001-8068-3806}\IEEEauthorrefmark{1}\IEEEauthorrefmark{2}%
    }
    \thanks{%
        \IEEEauthorrefmark{10} Both authors contributed equally to this research.
    }
    \IEEEauthorblockA{
        \textasteriskcentered~\textit{Integrated Systems Laboratory, ETH Zurich}, Switzerland \\
        \textdagger~\textit{Department of Electrical, Electronic, and Information Engineering, University of Bologna}, Italy \\
        \{phsauter,tbenz,paulsc,kgf,lbenini\}@iis.ee.ethz.ch
        }
    }
\fi

\maketitle

\begin{abstract}
Designing complex, multi-million-gate application-specific integrated circuits requires robust and mature electronic design automation (EDA) tools. 
We describe our efforts in enhancing the open-source Yosys+Openroad EDA flow to implement {\basilisk}, a fully open-source, Linux-booting RV64GC system-on-chip (SoC) design. 
We analyze the quality-of-results impact of our enhancements to 
synthesis tools, 
interfaces between EDA tools, 
logic optimization scripts, and 
a newly open-sourced library of optimized arithmetic macro-operators. 
We also introduce a streamlined physical design flow with an improved power grid and cell placement integration. 
Our {\basilisk} SoC design was taped out in IHP’s open \SI{130}{\nano\meter} technology. 
It achieves an operating frequency of \SI{77}{\MHz} (51 logic levels) under typical conditions, a \SI{2.3}{\x} improvement compared to the baseline open-source 
EDA flow, while also reducing logic area by \SI{1.6}{\x}. 
Furthermore, tool runtime was reduced by \SI{2.5}{\x}, and peak RAM usage decreased by \SI{2.9}{\x}. 
Through collaboration with EDA tool developers and domain experts, {\basilisk} establishes solid "proof of existence" for a fully open-source EDA flow used in designing a competitive multi-million-gate digital SoC.

\end{abstract}


\section{Introduction}

In recent years, interest in open-source \gls{eda} has significantly increased in academia and industry.
Academia benefits from collaboration free of \glspl{nda}, wide tool accessibility for students and researchers, transparent research on \gls{eda} tools, and free exchange of generated artifacts (e.g., netlists or layouts).
The \gls{eda} industry would benefit from an increased influx of skilled fresh talents trained by universities on \gls{eda} algorithms and their implementation in realistic open-source tool frameworks.
Meanwhile, the silicon design industry could benefit from reduced cost and, perhaps more importantly, sovereignty and a transparent chain of trust from \gls{rtl} descriptions to finished layouts.
As a consequence, \gls{oseda} tools have experienced a strong influx of users and developers, most prominently around the synthesis tool \emph{Yosys}~\cite{wolf2013yosys} and the \gls{pnr} tool \emph{OpenROAD}~\cite{ajayi2019openroad}. 

One key challenge in developing a strong open-source ecosystem is to raise the maturity and robustness of \gls{oseda} tools in handling large digital designs.
In this direction, Benz et al.~\cite{tbenz2023iguana} recently presented and released Iguana, a Linux-capable {\riscv} \gls{soc} design built on the configurable Cheshire \gls{soc} platform~\cite{ottaviano2023cheshire}.
Iguana combines an RV64GC core called CVA6, a HyperRAM DRAM controller, and a rich set of peripherals, including VGA and USB 1.1, to complete a representative, real-world Linux-capable system; its \gls{rtl} description is freely available~\cite{basilisk-git}.

Iguana's \SI{2}{\mega\gateeq}\footnote{Gate equivalent (\si{\gateeq}) is a technology-independent figure of merit measuring circuit complexity; it represents the area of a two-input, minimum-strength {NAND} gate.} implementation was first taped out in IHP's open \gls{pdk} 130nm technology~\cite{ihp-git} with a commercial closed-source tool flow~\cite{iguana-fsi23}.
Benz et al. open-sourced their work-in-progress \gls{oseda} flow (henceforth \emph{Iguana flow})~\cite{basilisk-git}, enabling others to build on their work.

In this work, we present {\basilisk}~\basiliskcite, the first end-to-end open-source Linux-capable \gls{soc} implemented in IHP's open \SI{130}{\nano\meter} technology from \gls{rtl} to tapeout.
Starting from the Iguana flow, we make systematic improvements to the involved \gls{eda} tools, flow scripts, and constraints to achieve a \gls{qor} that is not only acceptable for tapeout, but significantly exceeds the open-source state of the art.
{\basilisk} achieves an operating frequency of \SI{77}{\MHz}, a \SI{2.3}{\x} improvement over Iguana, while reducing the logic area by \SI{1.6}{\x} from \SI{1.8}{\mega\gateeq} to \SI{1.1}{\mega\gateeq}. 
We improve the runtime of synthesis by \SI{2.5}{\x} from \SI{5.4}{\hour} to \SI{2.2}{\hour} and the peak RAM usage by \SI{2.9}{\x} from \SI{217}{\giga\byte} to \SI{75}{\giga\byte}. 
During our tapeout, {\basilisk}'s \gls{pnr} completed with zero remaining \gls{drc} violations.

For the {\basilisk} \gls{soc} design, we update the Cheshire \gls{soc} platform to the newest version, adding new features, including a USB OHCI controller, to increase the capabilities and use cases.
We do not simplify the original \glsunset{rtl}\gls{rtl} description of the Cheshire \gls{soc} platform, which uses industry-grade {\sv} constructs, to avoid tool weaknesses, instead we focus on improving tools and the \gls{oseda} flow.
In the collaborative spirit of open source, we are collecting knowledge and leveraging existing efforts on cutting-edge algorithms and \gls{oseda} tools. 

We focus our efforts toward a \gls{qor}-optimized yet human-understandable tool flow; this \gls{hitl} philosophy allows the designers to understand and better evaluate each implementation step.
Finally, we provide the complete {\basilisk} \gls{soc} design and its critical, hard-to-implement parts, such as the \gls{fpu} or CVA6's scoreboard, as challenging benchmarks for all steps of synthesis and \gls{pnr}.
These benchmarks allow the community to push, optimize, and further improve \gls{oseda} tools and the corresponding flow scripts beyond the scope of this work. 

In this work, we present the following contributions:
\begin{itemize}
    \item An extensive study on the state-of-the-art open-source \gls{eda} flow using \emph{Yosys} for logic synthesis and \emph{OpenROAD} for \gls{pnr}. Starting from the Iguana flow, we identify \gls{qor} improvements in the flow steps with a particular focus on the synthesis engine and scripts.
    \item The integration of a library of hand-optimized implementations of arithmetic macro-operators in the Yosys-based synthesis flow and its open-source release~\pulpwarecite.
    \item A \gls{qor}-optimized \gls{hitl} open silicon implementation flow from {\rtgn} on an open-source foundry-supported \gls{pdk} qualified for manufacturing in regularly scheduled runs for both full reticle and \gls{mpw} runs~\cite{ihp-timetable}.
    \item {\basilisk}, the first end-to-end open-source Linux-capable \gls{asic} implemented in IHP's open \SI{130}{\nano\meter} node achieving \SI{77}{\MHz} with a logic area of \SI{1.1}{\mega\gateeq}.
\end{itemize}

\section{Related Work}
The OpenROAD~\cite{ajayi2019openroad} developers maintain an example flow called \gls{orfs}~\cite{orfs-git}. 
\gls{orfs} integrates a variety of open \glspl{pdk} (platforms) and a handful of small example designs. 
\gls{orfs} serves as a reference flow to get designers started, and as a benchmark to end-to-end verify OpenROAD on example designs. 
\gls{orfs} further provides automated design space exploration, facilitating the collection of key metrics using the given input parameters for each run.
With \gls{orfs}, all design-technology options are implemented using the same set of flow files configured through environment variables set by project-specific Makefile fragments. 
While providing a single solution that fits all design-technology options, such an approach makes the flow internals harder for non-expert tool developers to modify and tune.
The {OpenLANE}~\cite{ghazy2020openlane} flow provides a turnkey {\rtg} flow using Yosys and OpenROAD, similar to \gls{orfs}. 
It is maintained by {Efabless} and used in their popular Caravel~\cite{caravel-git} \gls{mpw} shuttles.
To complete the flow, OpenLane uses \emph{OpenRCX} to extract parasitics, \emph{Magic} to stream out the final fabrication files and perform {DRC} checks, and \emph{netgen} to complete {LVS} checks. 
A turnkey flow massively reduces the barrier of entry to designing \glspl{asic} by novice designers, but also makes it harder for expert designers to exercise the low-level tool control required to implement and optimize large and complex designs that push the capabilities of the \gls{oseda} tools to the limit.

\begin{figure}[htpb]
  \centering
  \includegraphics[width=\linewidth]{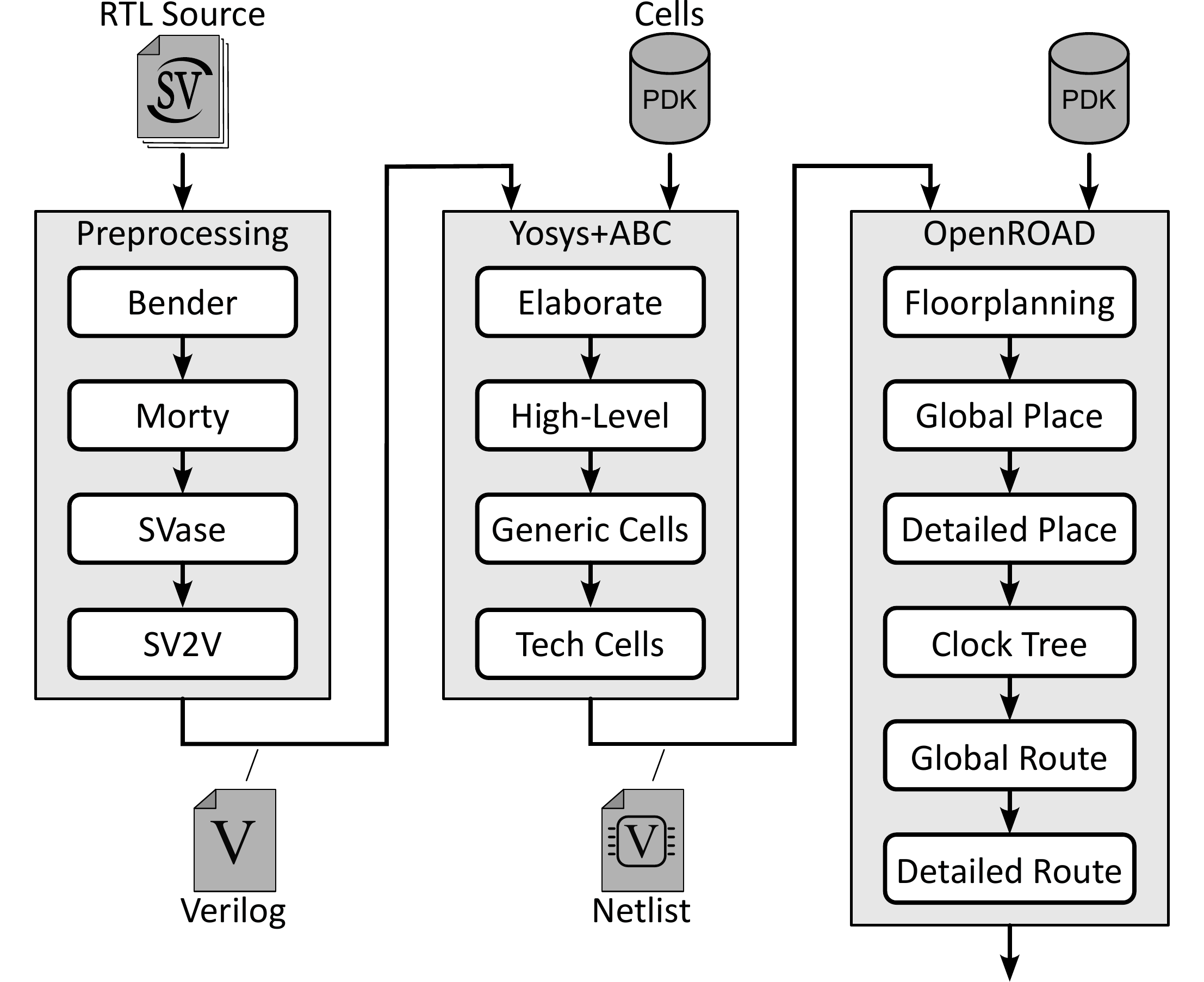}
  \caption{Preprocessing, synthesis and \gls{pnr} flow used for Iguana and \basilisk}    
  \label{fig:basilisk-flow} 
\end{figure}

\emph{Qflow}~\cite{edwards2021real} is one of the earliest complete \gls{oseda} {\rtg} toolchains using either \emph{VTR} (\emph{ODIN-II} and \emph{ABC~\cite{brayton2010abc}}) or Yosys to synthesize a Verilog design and implement the backend using their own placement and routing engines. 
Magic is again used to complete and check the final layout files.
The \emph{Raven \gls{soc}}~\cite{edwards2021real} is a working \gls{asic} designed with Qflow.
While able to implement a working \gls{soc}, this flow is currently limited to small designs under \SI{100}{\kilo\gateeq} and larger technology nodes in the range of \numrange{0.18}{0.5}\si{\micro\metre}.

\emph{ALLIANCE}~\cite{chaput2019risc} is another full toolchain, synthesizing small designs written in {VHDL} and implementing them in a portable \gls{cmos} technology. 
The \gls{pnr} flow of {ALLIANCE} has been replaced by \emph{CORIOLIS}~\cite{chaput2019risc} to support larger designs on the order of \SI{150}{\kilo\gateeq}. 
The developers recently started integrating Yosys into the {ALLIANCE/CORIOLIS} flow to synthesize designs. 
{ALLIANCE} and {CORIOLIS} use \emph{symbolic layouts} to implement the \gls{asic} backend, limiting their scaling to nodes above \SI{130}{\nano\metre}~\cite{chaput2019risc}. 
Compared to regular standard-cell-based \gls{pnr}, symbolic layout implements the design without any technology-specific design data. 

\emph{iEDA}~\cite{10473983} is a recent addition to the set of open \gls{pnr} tools proven to implement layouts in \SI{110}{\nano\metre} and \SI{28}{\nano\metre} technology nodes.
While being a complete \gls{pnr} framework, {iEDA} does not yet have an active community around its ecosystem and its documentation is written in Chinese with only a partial translation into English available. 

In addition to the above open {\rtg} flows, various standalone synthesis and \gls{pnr} tools and frameworks exist. 
The \emph{EPFL logic synthesis}~\cite{soeken2018epfl} libraries allow optimizations of structural netlists; they provide a standalone tool called \emph{mockturtle} built around these libraries, Yosys can be used to parse behavioral Verilog and convert the netlist to  a supported format.
\emph{LSOracle}~\cite{neto2019lsoracle} uses {EPFL logic synthesis} libraries to implement optimization passes on \glspl{aig} and is available as a Yosys plugin.
\emph{GHDL}~\cite{biagetti2023open} is a capable open-source simulator for the {VHDL} language with experimental support for converting {VHDL} designs to structural {VHDL} netlists;  
it is thus usable as a frontend to dedicated synthesis tools like Yosys. 
\emph{LibrEDA}~\cite{libreda} is a recently developed open framework facilitating the development of \gls{pnr} tools targeting research and education; it currently cannot produce implementable \gls{asic} layouts. 

Most users of these tools either target \glspl{fpga} or implement designs aimed at {Efabless'} {Caravel} \gls{mpw} shuttles in {SkyWater's} \SI{130}{\nano\metre} node.
{Caravel} designs are limited to a maximum user-defined core area of \SI{10}{\square\milli\metre} and implemented designs are usually under \SI{150}{\kilo\gateeq} with an average core density of \SIrange{10}{30}{\percent}~\cite{khan2023ghazi,zhu2022greenrio,khan2023ibtida}. 
{Caravel} designs remain well within the established capabilities of the \gls{oseda} tools and are unlikely to stress them. 
We aim to push \gls{oseda} beyond its current limits towards supporting end-to-end open implementation of large multi-million-gate designs, focusing on developing the tools and flows to remove their current \gls{qor} and runtime bottlenecks.

\section{Synthesis}

Yosys~\cite{wolf2013yosys} is a leading open-source synthesis engine widely used in \gls{oseda} flows. 
At the time of writing, it offers limited support for {\sv} language constructs.
With Cheshire's openly available \gls{rtl} description written in industry-grade {\sv}, significant preprocessing work is needed to convert the design's \gls{rtl} source to the simpler Verilog format that Yosys can parse. 
We use a chain of tools~\cite{tbenz2023iguana} to achieve this: %
\emph{Bender}~\cite{bender-git} collects and manages the dependencies from git repositories, \emph{Morty}~\cite{morty-git} combines all source files into a single compile context, \emph{SVase}~\cite{svase-git} propagates parameters and simplifies the most complex {\sv} constructs, and \emph{SV2V}~\cite{sv2v-git} converts the resulting simpler {\sv} code to behavioral Verilog.

At the first step of synthesis, Yosys parses this behavioral Verilog code into an \gls{ast}. 
Next, the syntax is elaborated into Yosys' main internal representation called \gls{rtlil} converting the behavioral code into a structural representation.
The structural \gls{rtlil} first uses high-level constructs such as arithmetic operations or \glspl{fsm} cells to represent the design.
Yosys executes commands, called passes, on the \gls{rtlil} description to progressively optimize and transform this high-level structural description into a low-level representation using only generic standard cells (MUX, AND, NOR, \dots).
Finally, Yosys maps sequential elements to provided technology cells and calls {ABC}~\cite{brayton2010abc} on the combinational networks to optimize the logic representation further and map the generic gates to specified technology cells.

To improve \gls{qor}, we optimize three aspects of the synthesis chain. 
The \gls{qor} of the Iguana baseline flow and our contributions are summarized in \Cref{tab:synth}. The reported results show cumulative improvements from left to right.
\begin{table}[t!]
    \centering
        \centering
        \caption{%
            Cumulative synthesis improvements from left to right; large impacts of each step are highlighted.%
        }%
        \label{tab:synth}
        \renewcommand*{\arraystretch}{0.95}
        \begin{threeparttable}
            \begin{tabular}{ccccc} \toprule
                &
                \textbf{Iguana~\cite{tbenz2023iguana}} &
                \textbf{MUX} &
                \textbf{ABC} &
                \textbf{LAU} \\

                \midrule

                \textit{Logic area} &
                \SI{1.8}{\mega\gateeq} &
                \textbf{\SI{1.4}{\mega\gateeq}} &
                \textbf{\SI{1.1}{\mega\gateeq}} &
                \SI{1.1}{\mega\gateeq} \\

                \textit{Timing} &
                \SI{33}{\MHz} &
                \SI{37}{\MHz} &
                \textbf{\SI{71}{\MHz}} &
                \textbf{\SI{77}{\MHz}} \\

                \textit{Logic levels~\tnote{a}} &
                182 LL &
                149 LL &
                \textbf{54 LL} &
                \textbf{51 LL} \\

                \textit{Runtime~\tnote{b}} &
                \SI{5.4}{\hour} &
                \textbf{\SI{2.8}{\hour}} &
                \SI{2.2}{\hour} &
                \SI{2.2}{\hour} \\

                \textit{Peak RAM~\tnote{c}} &
                \SI{217}{\giga\byte} &
                \textbf{\SI{105}{\giga\byte}} &
                \SI{76}{\giga\byte} &
                \SI{75}{\giga\byte} \\

                \bottomrule
                
            \end{tabular}

            \begin{tablenotes}[para, flushleft]
                \item[a] Number of logic gates in longest path
                \item[b] \SI{2.5}{\GHz} Xeon E5-2670
            \end{tablenotes}
        \end{threeparttable}
\end{table}

\subsection{Part-Select Synthesis  (MUX)}
\label{synth:mux}

\begin{figure}[htpb]
  \centering
  \includegraphics[width=\linewidth]{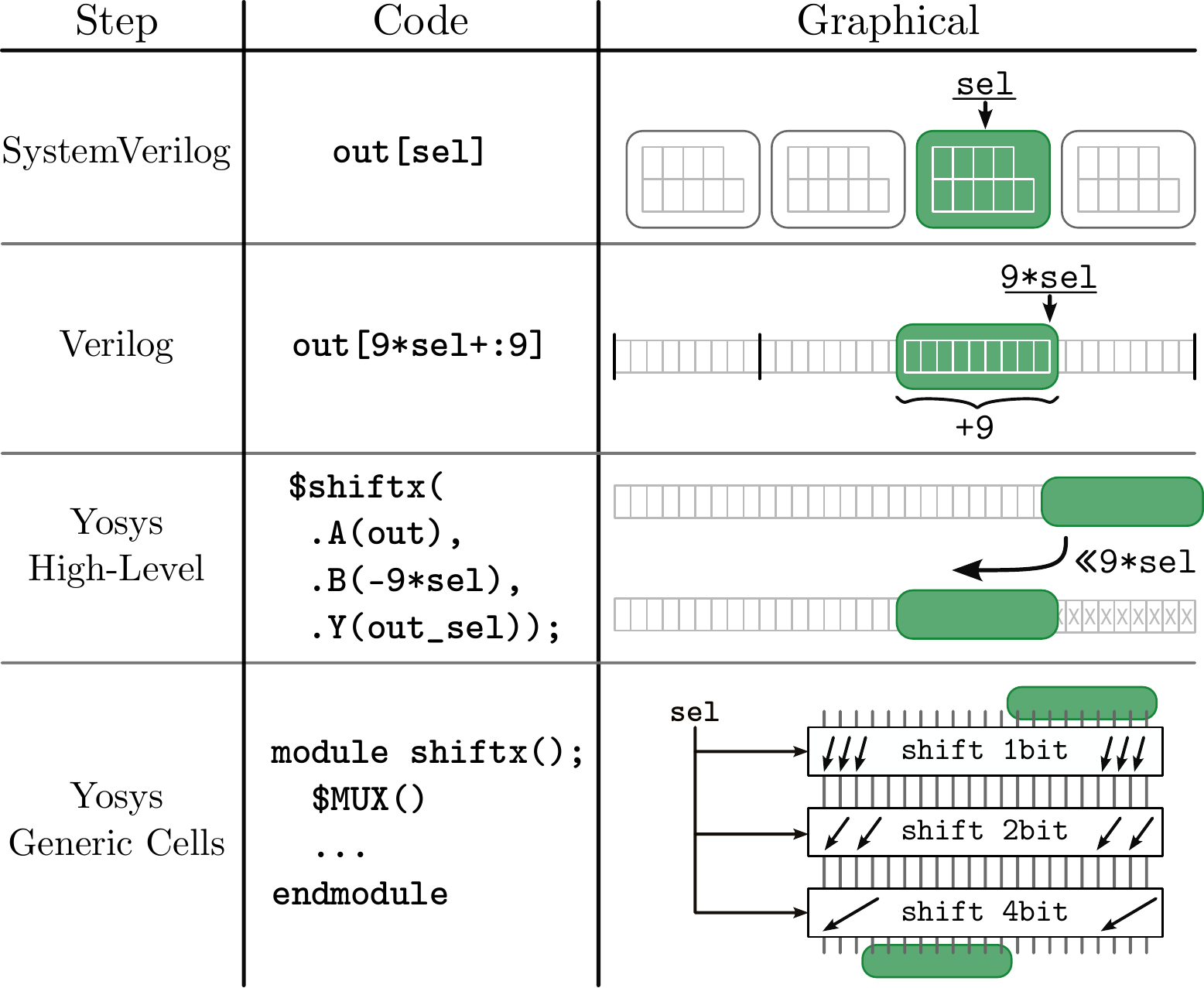}
  \caption{Step-by-step preprocessing and synthesis of a part select operation}    
  \label{fig:part-select} 
\end{figure}

Yosys versions before our improvements\yosysthanksnote (\emph{$<$0.34}) elaborate indexed part-select operations to shift operations instead of more efficient \emph{block-multiplexer trees}.
A step-by-step overview of the previous \emph{shift operation} inference is given in \Cref{fig:part-select}.

First, \emph{SV2V} preprocesses the {\sv} representation describing multi-dimensional arrays to a one-dimensional packed array with the size of the respective inner dimension as its stride. 
Then, Yosys elaborates the \gls{rtl} description to the high-level \gls{rtlil} cell \emph{\$shiftx}, representing a shift where vacated positions are filled with don't-care bits.
Shifting the selection/readout window is logically equivalent to directly selecting a group of bits.
Yosys chooses \emph{shift-by-N barrel shifters} to implement block selection prior to our fix.
By not limiting the shift magnitude to a multiple of the stride, this approach produces redundant hardware capable of handling \emph{invalid and unused} block selection scenarios.
To support these unnecessary selections incurs significantly more area at a longer critical path than a simple \emph{block multiplexer}.

This causes Yosys to implement a \emph{MUX} tree many times larger and more complex than necessary. 
Constraining the magnitude of the shift in the low-level representation results in a significant amount of unused logic and many don't-care bits. 
The optimizations implemented in Yosys after this mapping process are not powerful enough to remove all unnecessary and redundant parts of the shifter, further propagating this complex logic through the flow.
By the time the netlist is given to {ABC}, all remaining don't-care bits have been converted to logic zeros, blocking potential optimizations.
This implementation of part selects using general barrel shifters thus significantly inflates design area and the number of logic levels in part-select paths.
Due to the inflated logic description, peak {RAM} usage and synthesis time also increase by \SI{2}{\x} (see \Cref{tab:synth}).

Instead of elaborating part selects to block multiplexers, we developed an additional optimization pass reducing all eligible shift operations in the design to block multiplexers. 
To achieve this, we detect constant strides in the control logic of a shift operation describing behavior equivalent to a part select. 
The pass then increases the stride to the next power-of-two value with the input and output padded accordingly. 
Increasing the stride to a power of two causes all lower-weighted barrel-shift stages to receive a well-defined constant input, allowing for trivial optimization using constant propagation. 
The additional padded bits are optimized away using existing Yosys passes.

An alternative implementation would be to directly infer block multiplexers from part select operations at the elaboration stage.
We evaluate our optimization-pass-based solution against this more direct approach.
In isolated benchmarks on part select operations, this direct solution produces the same hardware implementation as our high-level optimization pass.
On an entire design, our optimization pass can optimize additional shift operations and thus produce a better overall design implementation than the simpler direct method.

As can be seen in \Cref{tab:synth}, our optimization pass reduces the logic area by \SI{22}{\percent} and increases the operating frequency by \SI{12}{\percent}. 
The computing resource utilization during synthesis is significantly reduced, the peak {RAM} usage is \SI{52}{\percent} lower, and the synthesis runtime is \SI{48}{\percent} lower.
\subsection{ABC Scripts Overhaul (ABC)}
\label{synth:abc}

\begin{figure}[htpb]
  \centering
  \includegraphics[width=\linewidth]{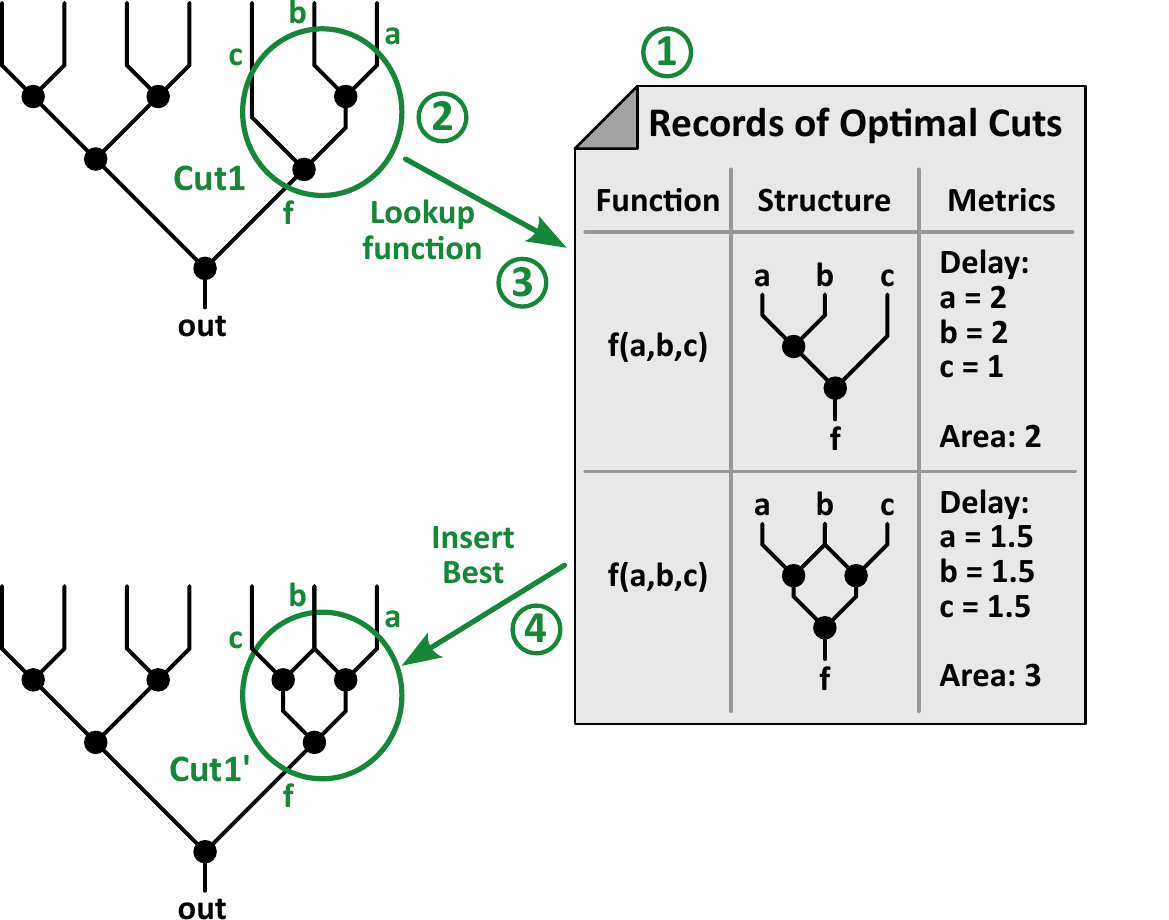}
  \caption{A \gls{lms} example for a simple example network (left). 
  Three input cuts are generated from the network and the three input records (right) are probed. 
  The best implementation is taken and inserted into the graph.}
  \label{fig:lazy-mans-synthesis} 
\end{figure}

In cooperation with the community of researchers and practitioners developing and using {ABC}\abcthanksnote, we overhaul the Yosys-internal ABC script for logic optimization and technology mapping.
We leverage \gls{lms}, as proposed by Yang et al.~\cite{lazy-synthesis}, to improve the \gls{qor} while keeping the impact on runtime minimal.
\Gls{lms} uses a pre-generated library of optimal logic structures to replace so-called \emph{cuts}, logic blocks of equal input-output behavior in the netlist.
An overview of \gls{lms} is shown in \Cref{fig:lazy-mans-synthesis} for a simple example network (left).
A library of records \circled{1} is generated in advance. 
The netlist is divided into cuts \circled{2} with the same number of inputs as the records.
Then the library of records is probed \circled{3} for structures implementing the same logic function. Finally, an optimal structure is selected and inserted back \circled{4} into the netlist.

Generating this library of optimal logic structures is a time-consuming process, but is highly justifiable as its generation is only required once. 
The cuts (records) are derived by running other optimization techniques on a large variety of different benchmarks and saving superior implementations of any logic function. 
Specifically, the record is created from six input cuts; e.g., each record can be mapped to one six-input \gls{lut}.
Records also contain the subsets of cuts with fewer inputs. 
To improve the library quality, we extend the 6-input record using cuts obtained from {\basilisk} through the \lstinline{rec_add3} command available in {ABC}.

Once the record of cuts is loaded into ABC (\lstinline{rec_start3}), the \gls{lms} optimizer can be called via \lstinline{&if -y -K NUM}, where \lstinline{NUM} is the number of inputs of the cuts stored in the record. 
The \gls{lms} flow presented in this work is closely based on the flow as presented by Yang et al.:
\begin{lstlisting}[language=abc]
    &st; &if -y -K 6; &syn2; &b; 
    &st; &dch -x; &if -K 4;
\end{lstlisting}
This iteration is applied multiple times; ten to twenty iterations are required to converge on a near-optimal solution.
The script first performs structural hashing (\lstinline{&st}) followed by \gls{lms} integrated into the mapper (\lstinline{&if}) command.
\lstinline{&syn2} and \mbox{\lstinline{&b}} rewrite and balance the depth of the graph, respectively. 
Structural choices are computed (\lstinline{&dch}) and then considered when mapping to four-input~\glspl{lut}. 
In the last step, it is possible to map to any type of \gls{lut}.
We find that selecting four-input \glspl{lut} produces the best results for {\basilisk}.
Finally, the ABCs technology mapper \lstinline{&nf} maps the netlist to standard cells.
\begin{lstlisting}[language=abc]
    &st; &nf -D 6000;
\end{lstlisting}
The mapping to standard cells is executed strictly after completing all optimization iterations.
If the mapping iteration is used before the optimization converges on a solution, the overall \gls{qor} is reduced.
We experiment with adding structural choices before the technology mapper but see only very marginal \gls{qor} gains at a high runtime cost.

In addition to using \gls{lms}, we extend the \lstinline[language=yosys]{abc} command in Yosys to allow for more control over how the library and netlist are loaded into {ABC}.
This exposes the \lstinline{read_lib} command used to load the standard cell library, making it possible to define the additional parameters \emph{-S slew -G gain}.
These parameters are required for {ABC} to generate delays for each cell derived from the liberty timings. 
Previously, Yosys would not set these values, causing {ABC} to completely ignore the liberty timings and use unit delays instead. 
Using the properly loaded timing model increases the \gls{qor} obtained from \lstinline{&nf}.
Using a parametric sweep on a subset of {\basilisk} modules, we find \emph{-S 20 -G 3} to be suitable parameters for IHPs \SI{130}{\nano\metre} technology node.

The improved {ABC} script based on \gls{lms}, together with the correct delay models, improves the \gls{qor} substantially. The area is \SI{21}{\percent}  smaller and the critical path improves by \SI{1.9}{\x} with significantly shorter runtime and reduced peak RAM usage, see \Cref{tab:synth}.

\subsection{Library of Arithmetic Units (LAU)}
The critical path of the {\basilisk} design is in the datapath of its \gls{fpu};  
it traverses a 53$\times$\SI{53}{\bit} multiplier followed by two additions ($y=a\times b +c +d$), the latter with a 163-\si{\bit} integer.
We manually implement this datapath using modules from our library of arithmetic units, detail the necessary changes to Yosys, and present our approach's improved \gls{qor}.

Yosys maps arithmetic units to generic standard cells using built-in implementations of high-level blocks loaded from Verilog descriptions.
The same fixed implementation is used for all instances of the same type without consideration of timing and area constraints. 
For example, all adders are implemented using a \gls{ppa-bk} architecture without any alternatives available.

Some complex operations have specialized passes to infer and implement them. 
Most importantly, a generalized sum-of-products ($y=a\times b + c\times d + 1\times e\dots$) can be inferred from the current \gls{rtlil} using the Yosys pass \lstinline[language=yosys]{alumacc} which is later on implemented in the \lstinline[language=yosys]{maccmap} pass. 
Additionally, Yosys has a \lstinline[language=yosys]{booth} pass to implement multipliers. 
Since \lstinline[language=yosys]{maccmap} combines multipliers and adders into multiply-accumulate cells and \lstinline[language=yosys]{booth} transforms multipliers into generic standard cells, a designer needs to choose which approach they prefer; it is not possible to combine \lstinline[language=yosys]{booth} and \lstinline[language=yosys]{maccmap} without code changes to  architectures implemented in {C++} using Yosys' internal functions. 
Yosys' current way of implementing arithmetic operations makes it thus difficult to add more arithmetic operations and support additional architectural choices, targeting different scenarios.

Building on the work by R. Zimmermann~\cite{zimmermann1998library}, we build a library of arithmetic units to supersede Yosys' existing mappings. 
Our open-source library~{\pulpwarecite} has three speed grades optimized for different design points for each arithmetic operation.
Yosys has an existing pass to match sub-circuits and wrap them into a custom cell called \lstinline[language=yosys]{extract}. 
We extend the \lstinline[language=yosys]{extract} pass to support variable width operators to match arbitrary arithmetic operations. 
We can improve area, timing, and power by extracting functionally equivalent sub-circuits and then implementing them with the custom mapping from our library of arithmetic units.
To demonstrate the effectiveness of this approach, we create an optimized mapping for the critical data path, which is usable as part of the Yosys synthesis script, and benchmark it against the existing approaches in Yosys.

\begin{figure}[t]
  \centering
  \includegraphics[width=\linewidth]{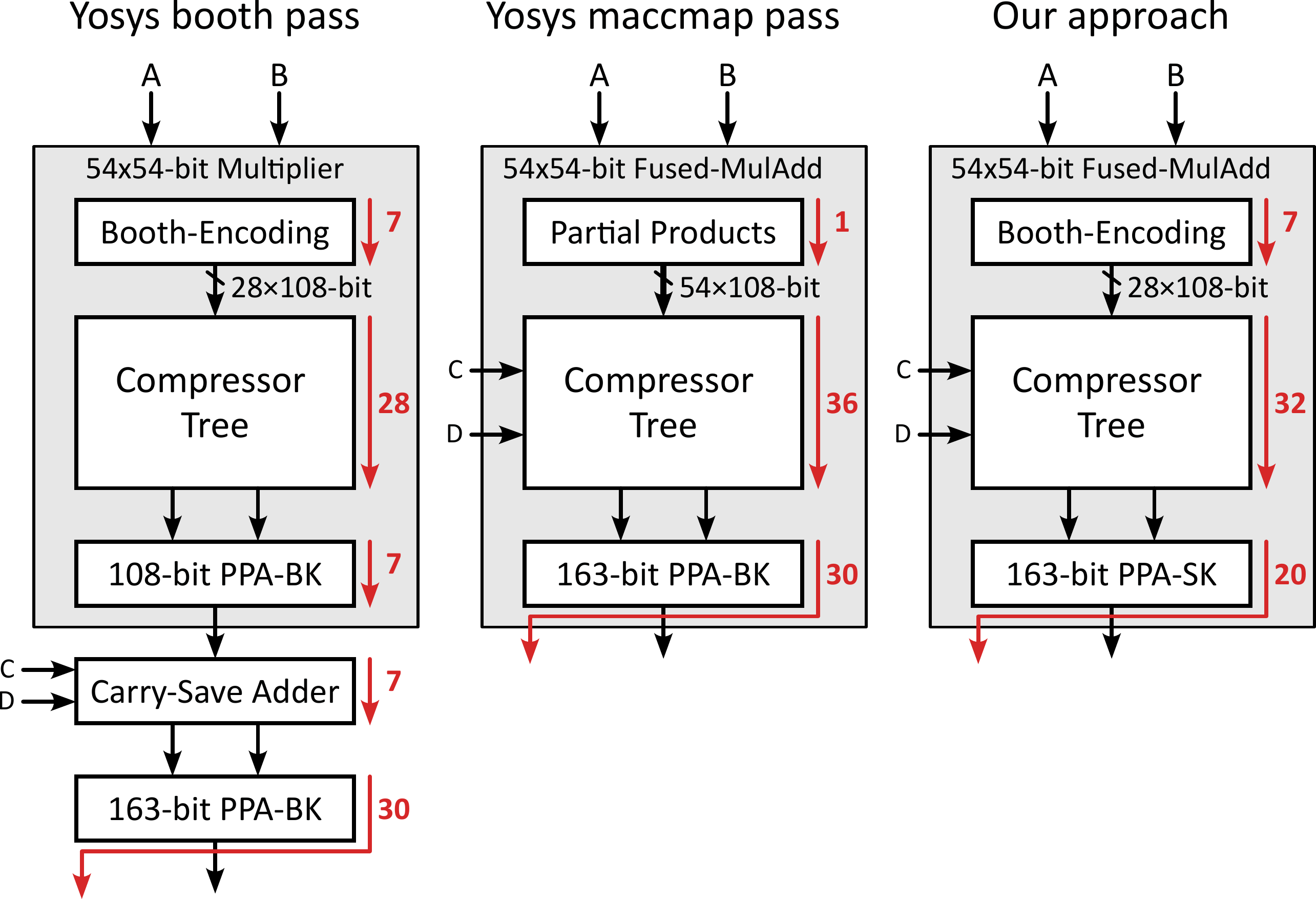}
  \caption{Yosys elaborated and synthesized datapath (left, middle) and our optimized datapath (right) with unit delays.}    
  \label{fig:fma} 
\end{figure}

\Cref{fig:fma} compares the architectures implemented by \mbox{\lstinline[language=yosys]{booth},} \lstinline[language=yosys]{maccmap}, and our implementation using the library of arithmetic units. 
The theoretical unit-gate delays for each block are given, as they do not consider driving strength, wires, and other effects; they should be used to get an overview of the rough cost of each operation, not for accurate comparisons.
\lstinline[language=yosys]{booth} (left) implements the multiplication using a radix-4 Booth encoder followed by a \emph{Wallace tree} compressor.
The compressor has a lower depth since the Booth encoding reduced the number of partial products from 54 to 28. 
The final \gls{cpa} for the multiplication is implemented using a \gls{ppa-bk}. 
The summation is performed using a \gls{csa} followed by the final \gls{ppa-bk}. 
This architecture is constrained due to the separation of the partial product summation and the final summation into two discrete steps. 
The \lstinline[language=yosys]{maccmap} implementation (middle) improves on this by fusing the additions into the compressor tree implementing partial product summation. 
It does not employ Booth encoding to reduce the number of partial products, creating the deepest compressor tree. 
The final \gls{cpa} is also a \gls{ppa-bk} architecture. 

Our implementation (right), built from modules in the library of arithmetic units, uses Booth-encoding and the faster \gls{ppa-sk} \gls{cpa} architecture to reduce the critical path.

The literature on the efficacy of radix-4 Booth encoding is split. Wolfgang et al.~\cite{wolfgang2002Booth} formalize an analytical model and show a delay advantage in favor of Booth encoding. 
Shahzad et al.~\cite{shazad2015Booth} implement a Booth-encoded and non-encoded multiplier using a reduced complexity Wallace tree and show an increase in delay using Booth encoding for most multiplier widths. 
We implement both variants and find the radix-4 Booth-Wallace architecture to produce superior \gls{qor}.
Currently, all approaches map to generic standard cells and are passed to {ABC}, which may change the architecture. 
Still, passing an improved architecture to {ABC} is likely to produce a better final netlist.

Compared to the \lstinline[language=yosys]{booth} implementation used before, our approach reduces the critical path by \SI{9}{\percent}, while it is reduced by \SI{11}{\percent} compared to the alternative Yosys flow using \lstinline[language=yosys]{maccmap}, see \Cref{tab:synth}.

\section{Place \& Route}
We use \emph{OpenROAD}~\cite{ajayi2019openroad} to implement {\basilisk}'s synthesized netlist. 
As a reference, we re-run the Iguana flow; the final result has hundreds of \gls{drc} violations left after detailed routing. 
Using a commercial logic synthesizer, we can fix problematic modules such as the boot ROM, CVA6's scoreboard in the issue stage, and the logic surrounding caches to enable proper comparisons.
Analyzing Iguana's \gls{pnr}, we identify improvement steps mainly in the \emph{\gls{eda} tool flow} (how the individual components of {OpenROAD} are invoked) and the physical constraints of the \gls{asic}.

Top metal power routing needs to pass through all metal layers to connect to the power rails of the standard cells. 
The created via stacks locally block routing resources.
We improve the routability of the design by reducing the width of each power stripe. 
We increase their count on the top metal layer to compensate for the loss in current carrying capacity. 
Overall, this creates a more homogeneous pattern for \gls{pnr} and directly eases local routing congestion underneath the stripes by reducing the number of blocked resources per stripe.

We change the SRAM configuration of the caches to reduce the number of SRAM macros from 36 to 24 by using larger SRAMs and adjusting the cache organization without changing the total size.
We arrange the SRAM macros to maximize the uninterrupted core area and minimize routing channels. 
This reduces routing congestion as the macros block most metal layers, making routing resources above them sparse.

Currently, OpenROAD cannot restructure the netlist to improve routability or the timing of the physical implementation.
It only inserts or removes buffers and resizes cells. 
This places additional pressure on the synthesizer as the netlist needs to be already routable. 
Yosys does not consider timing, wire fanout, or placement and will thus reduce any redundant logic, creating very high fanout nets.
{ABC}, on the other hand, limits the maximum fanout and considers the standard cell library timings but otherwise does not consider placement and routing effects during optimization or mapping.
These effects combined make it often difficult or impossible to implement netlists of complex designs without carefully configuring the design and the synthesis parameters. 

Very dense modules with random routing patterns, such as boot ROMs, are a particular sources of issues. 
OpenROADs global placement engine does support a routability-driven mode, where dense routing regions are identified and the apparent size of connected cells is inflated.
The routability-driven mode artificially increases the calculated cell density in global placement and lowers the actual cell density.
This, in turn, lowers congestion by increasing the routing resources available per cell.

The default \emph{hyper-parameters} of the global placement stage do not trigger the built-in \emph{routability-driven} mode.
The global placer will instead produce homogeneous placement density and the lowest possible wire length, not considering any local routing pattern, which may cause congestion.
As OpenROAD currently only accepts global (as opposed to region- or instance-based) settings, we tune several \emph{hyper-parameters} of the global placement engine to improve the placement of dense blocks.
With proper tuning, we can get a routable design without any \gls{drc} violations.
Specifically, we start routability-driven placement earlier by increasing \lstinline[language=other]{routability_check_overflow} and use a larger cell-area inflation ratio to more heavily penalize cells causing congestion by increasing both \lstinline[language=other]{routability_inflation_ratio_coef} and \lstinline[language=other]{routability_max_inflation_ratio}.

We observe OpenROAD's \emph{global routing} to over-prioritize the lowest metal layers when planning the routing process.
This increases the congestion close to the standard cells, making it difficult for detailed routing to fix remaining violations as it cannot find a clear path from the higher levels down to the lower metals in already congested regions. 
We reduce the \emph{target metal utilization} of the lowest two metals by \SI{30}{\percent} using \lstinline[language=other]{set_global_routing_layer_adjustment} to push global routing more onto the higher metal layers, increasing the flexibility for routing changes in the detailed router.

\begin{figure}[ht!]
    \centering
    \subfloat[\label{fig:die:iguana}]{{\includegraphics[width=\linewidth]{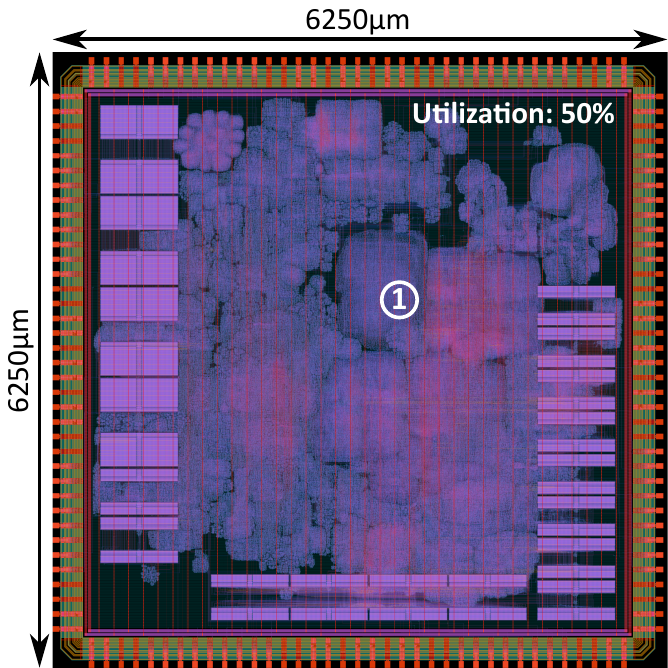}}}%
    
    \subfloat[\label{fig:die:basilisk}]{{\includegraphics[width=\linewidth]{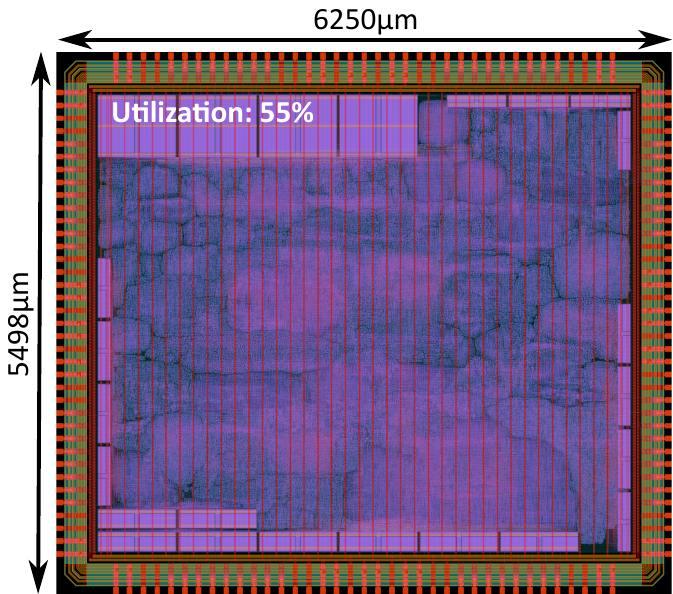}}}%
    \caption{Layout files produced by running the original Iguana flow (a) and of {\basilisk} (b).}    
    \label{fig:die-comparison} 
\end{figure}

\Cref{fig:die-comparison} shows the die shot of our Iguana baseline flow (a) and the newest version of the {\basilisk} design (b).
{\basilisk} has a more spread-out placement with less-distinct amoebae-shaped individual modules.
The global nature of the routability-driven hyper-parameters makes it difficult to decrease local placement densities without affecting the overall placement solution. 
This increases the total wire length but not to the degree that it noticeably affects the critical path and reduces the operating frequency. 
The benefit of our improvements is also visible; Iguana has to use more routing on the top metal layers in dense regions, creating red-tinted vertical lines in the image.
Requiring top metal routing indicates high local congestion in affected regions, as OpenROAD would otherwise prefer lower metal layers.
A prominent example of this effect is to the right of \circled{1} in \Cref{fig:die:iguana}. 
\begin{table}[t!]
    \centering
        \centering
        \caption{%
            Key metrics of {\basilisk}%
        }%
        \label{tab:basilisk}
        \begin{threeparttable}
            \begin{tabular}{cc} \toprule

                \textit{Logic area (NAND2)} &
                \SI{1.1}{\mega\gateeq} \\

                \textit{Logic levels~\tnote{a}} &
                51 LL \\

                \textit{Technology} &
                \SI{130}{\nm} IHP \\

                \textit{Operating frequency} &
                \SI{77}{\MHz} \\

                \textit{SRAM memory} &
                \SI{172}{\kibi\byte} (24 macros) \\

                \textit{Chip / core area} &
                \SI{39}{\mm} / \SI{21}{\mm} \\

                \textit{IO count} &
                68 \\

                \bottomrule
                
            \end{tabular}

            \begin{tablenotes}[para, flushleft]
                \item[a] Number of logic gates in longest path
            \end{tablenotes}
        \end{threeparttable}
\end{table}

\section{Results}
\begin{figure}[htb]
  \centering
  \includegraphics[width=\linewidth]{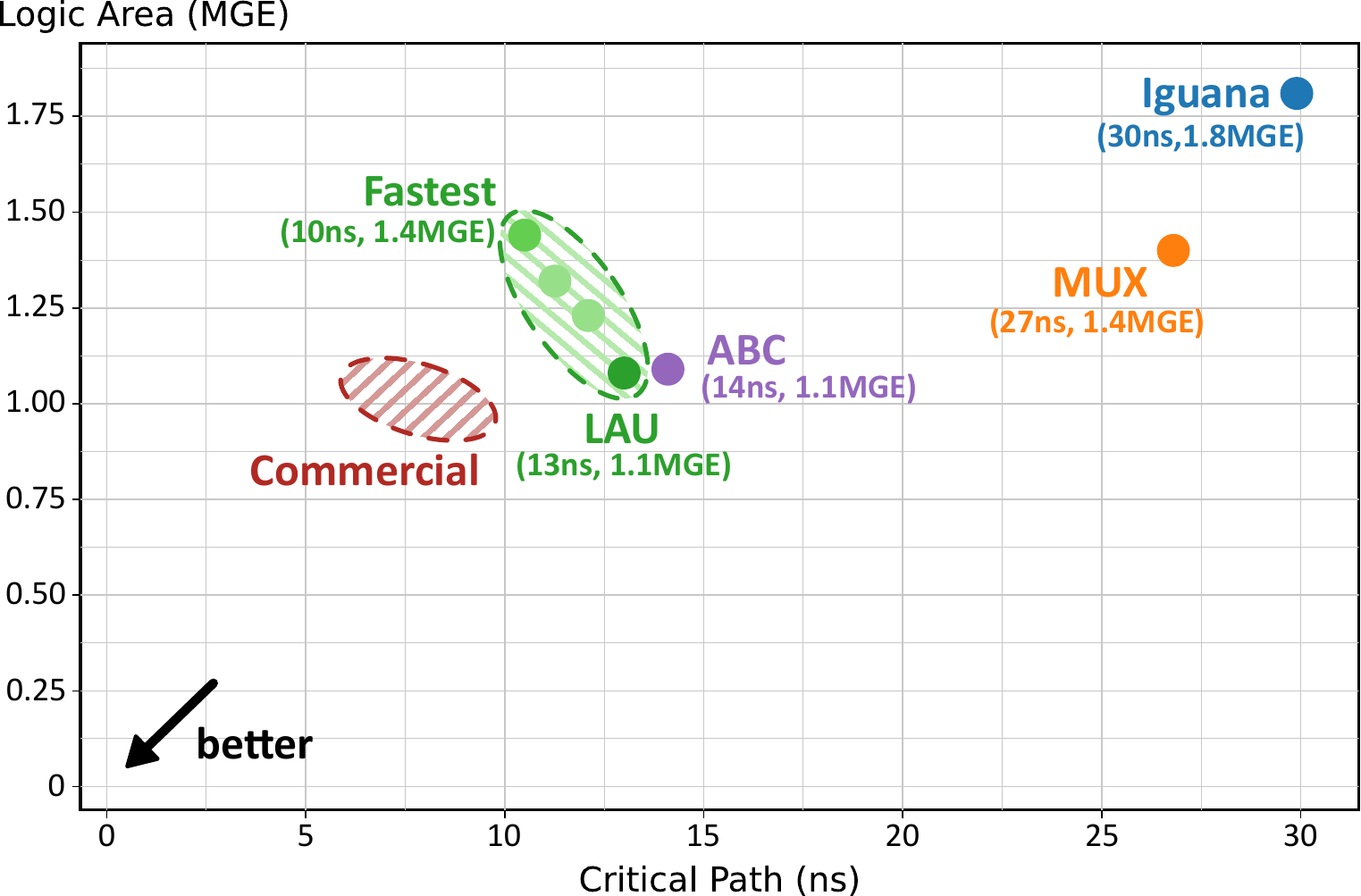}
  \caption{Synthesis and \gls{pnr} results of Iguana, our contributions and a commercial reference flow.}    
  \label{fig:at-plot} 
\end{figure}

The cumulative \gls{qor} improvements of our {\basilisk} design over the Iguana flow are shown in \Cref{tab:synth} and \Cref{fig:at-plot}.
We time the netlists obtained from Yosys in a commercial synthesis tool to ensure accurate timing reports using typical operating conditions.

In the {AT-plot}, Iguana denotes the baseline re-run of the Iguana flow presented by Benz et al.~\cite{tbenz2023iguana} with a logic area of \SI{1.8}{\mega\gateeq} and a critical path of \SI{30}{\ns}. 
Our first contribution (\emph{MUX}) to the synthesis of part-selects improves the logic area by \SI{22}{\percent} and the critical path by \SI{11}{\percent}.
Building an optimized ABC script used in Yosys, utilizing \gls{lms}, and providing parameters necessary to create accurate delay models show the largest \gls{qor} improvements (\emph{ABC}). 
The area is further reduced by \SI{21}{\percent} (\SI{39}{\percent} compared to the Iguana flow) and the critical path is lowered by \SI{2.1}{\x} to \SI{14.1}{\ns}.
Finally, the new approach to mapping high-level Yosys cells using our library of arithmetic units (\emph{LAU}) can further improve the critical path by \SI{9}{\percent} to \SI{13}{\ns}. 

The tight control over the optimization flow, gained through our optimized {ABC} scripts, and the multitude of choices when selecting architectural implementations of individual arithmetic operations allows our implementations to span a large area of the {AT-plot}.
This enables developers more flexibility when considering area-timing tradeoffs during the design and implementation phases.

The Pareto-optimal point improves the timing by \SI{2.3}{\x} and reduces the area by \SI{1.6}{\x} compared to the Iguana flow.
Using a timing-optimized variation of our ABC script, we can further decrease the critical path to \SI{10.4}{\ns} at the cost of an increased logic area.

Our contributions take a significant step towards closing the \gls{qor} gap of open-source flows compared to their commercial counterparts.
Still, commercial \gls{eda} tools have a clear edge on multi-million-gate designs like {\basilisk}. 
They achieve this with timing-aware synthesis, tighter integration of elaboration and optimizations, deeper libraries of pre-optimized blocks, and a stronger focus on backend-aware synthesis.
The best-achieved logic area and critical path length are within \SI{50}{\percent} of a commercial synthesis flow of {\basilisk}. 

\section{Conclusion and Outlook}

In this work, we evaluate the state-of-the-art open-source \gls{eda} flows and contribute significant improvements to the tools and flow, resulting in a flow viable for multi-million-gate \gls{soc} design tapeouts. 
In the process, we open-sourced a library of hand-optimized architectures of common arithmetic operations, which is compatible with Yosys.

Synthesizing and implementing {\basilisk} in IHP's open \SI{130}{\nm} technology, we optimize the design's clock frequency by \SI{2.3}{\x}~from \SI{33}{\MHz} to \SI{77}{\MHz} compared to the Iguana flow, while reducing the logic area from \SI{1.8}{\mega\gateeq} to \SI{1.1}{\mega\gateeq} and decreasing the synthesis runtime from \SI{5.4}{\hour} to \SI{2.2}{\hour}. 
Additionally, a timing-optimized synthesis script achieves a maximum frequency of \SI{97}{\MHz}.

Our improved \gls{pnr} scripts and constraints successfully implement the {\basilisk} \gls{soc} design with zero \gls{drc} violations with an increased core utilization of \SI{55}{\percent} compared to \SI{50}{\percent} using the Iguana flow.
All our improvements allowed us to tape out {\basilisk} successfully.

We contribute to improving open-source \gls{eda} tools by reporting and fixing tool issues, releasing our optimized \emph{flow scripts}, and implementing {\basilisk}, which can be used as a benchmark of a real-world multi-million-gate \gls{soc} design across the entire \gls{oseda} toolchain.

To fully utilize the library of arithmetic units, more effort is required to adjust the existing Yosys ABC9~\cite{barzen2023abc9} system for standard cell designs. 
Leveraging the white-box approach, the implementation of arithmetic units could be controlled from Yosys without losing proper delay calculation and logic optimization of any connected circuits in ABC. 
Further, the timing reported from ABC can be utilized to change the speed grade and implementation during synthesis, allowing for the automatic selection of the optimal architectures for each individual arithmetic operator.

Work towards an integrated timing- and constraints-driven synthesis and \gls{pnr} flow is required in a long-term effort.
This would enable more aggressive timing, area, and routability optimizations where needed. 
This requires a high degree of coordination and long-term planning between \gls{oseda} developers. 
Suitable standardized formats could facilitate this while maintaining the independence of tools and maintaining the ability to swap or replace individual tools in the flow, easing maintainability and integration of new research.

\section*{Acknowledgement}
\ifx\blind\undefined
    We thank %
    Alan Mishchenko, %
    Masahiro Fujita, %
    Giovanni De Micheli, %
    Andrea Costamagna, %
    Alessandro Tempia Calvino, %
    Osama Hammad Abdel Reheem, %
    Matt Liberty, %
    Martin Povišer, %
    the Yosys team, %
    Beat Muheim, %
    and %
    Zerun Jiang, %
    for their valuable contributions to the research project. %
    We further thank all contributors to the \gls{oseda} tools. %
    
    We are deeply grateful to IHP for their generous support and providing us with the opportunity for an open-source tapeout of this scale.

    This work was supported in part through the TRISTAN (101095947) project that received funding from the HORIZON CHIPS-JU programme.

\else
    \textit{Acknowledgments omitted for blind review.}
\fi



\begin{thebibliography}{10}
\providecommand{\url}[1]{#1}
\csname url@samestyle\endcsname
\providecommand{\newblock}{\relax}
\providecommand{\bibinfo}[2]{#2}
\providecommand{\BIBentrySTDinterwordspacing}{\spaceskip=0pt\relax}
\providecommand{\BIBentryALTinterwordstretchfactor}{4}
\providecommand{\BIBentryALTinterwordspacing}{\spaceskip=\fontdimen2\font plus
\BIBentryALTinterwordstretchfactor\fontdimen3\font minus
  \fontdimen4\font\relax}
\providecommand{\BIBforeignlanguage}[2]{{%
\expandafter\ifx\csname l@#1\endcsname\relax
\typeout{** WARNING: IEEEtran.bst: No hyphenation pattern has been}%
\typeout{** loaded for the language `#1'. Using the pattern for}%
\typeout{** the default language instead.}%
\else
\language=\csname l@#1\endcsname
\fi
#2}}
\providecommand{\BIBdecl}{\relax}
\BIBdecl

\bibitem{wolf2013yosys}
C.~W. et~al., ``Yosys - a free verilog synthesis suite,'' 2013.

\bibitem{ajayi2019openroad}
T.~Ajayi, D.~Blaauw, T.~Chan, C.~Cheng, V.~Chhabria, D.~Choo, M.~Coltella,
  S.~Dobre, R.~Dreslinski, M.~Foga{\c{c}}a \emph{et~al.}, ``{OpenROAD}: Toward
  a self-driving, open-source digital layout implementation tool chain,''
  \emph{Proc. GOMACTECH}, 2019.

\bibitem{tbenz2023iguana}
T.~Benz, P.~Scheffler, J.~Schönleber, and L.~Benini, ``Iguana: An end-to-end
  open-source {Linux-capable} {RISC-V SoC} in 130nm {CMOS},'' 2023.

\bibitem{ottaviano2023cheshire}
A.~Ottaviano, T.~Benz, P.~Scheffler, and L.~Benini, ``Cheshire: A lightweight,
  {Linux-Capable} {RISC-V} host platform for domain-specific accelerator
  plug-in,'' \emph{IEEE TCAS II}, 2023.

\bibitem{basilisk-git}
{PULP Platform contributors}, ``Iguana,''
  \url{https://github.com/pulp-platform/iguana}, 2024.

\bibitem{ihp-git}
{IHP-GmbH}, ``{IHP Open Source PDK},''
  \url{https://github.com/IHP-GmbH/IHP-Open-PDK}, 2022.

\bibitem{iguana-fsi23}
T.~Benz, P.~Scheffler, J.~Schoenleber, and P.~Sauter, ``Industry-grade
  {SystemVerilog} {IPs} and the open flow: How we synthesized {Iguana},''
  available:
  \url{https://wiki.f-si.org/index.php?title=Industry-Grade_SystemVerilog_IPs_And_The_Open_Flow:_How_We_Synthesized_Iguana}.

\bibitem{pulpware-git}
{PULP Platform contributors}, ``Pulpware,''
  \url{https://github.com/pulp-platform/pulpware}, 2024.

\bibitem{ihp-timetable}
{IHP - Leibniz Institute for High Performance Microelectronics}, ``Mpw schedule
  2024 \& 2025 and price information 2024,''
  \url{https://www.ihp-microelectronics.com/services/research-and-prototyping-service/mpw-prototyping-service/schedule-price-list}.

\bibitem{orfs-git}
{The-OpenROAD-Project}, ``{OpenROAD Flow},''
  \url{https://github.com/pulp-platform/OpenROAD-flow-scripts}, 2024.

\bibitem{ghazy2020openlane}
A.~Ghazy and M.~Shalan, ``Openlane: The open-source digital {ASIC}
  implementation flow,'' in \emph{WOSET}, 2020.

\bibitem{caravel-git}
{Efabless}, ``Caravel harness,'' \url{https://github.com/efabless/caravel},
  2022.

\bibitem{edwards2021real}
R.~T. Edwards, M.~Shalan, and M.~Kassem, ``Real silicon using open-source
  {EDA},'' \emph{IEEE Design \& Test}, 2021.

\bibitem{brayton2010abc}
R.~Brayton and A.~Mishchenko, ``{ABC}: {An} academic industrial-strength
  verification tool,'' in \emph{CAV}, 2010.

\bibitem{chaput2019risc}
J.-P. Chaput, M.-M. Lou{\"e}rat, R.~Chotin-Avot, and A.~Satin, ``{RISC-V}
  design using free open source software,'' in \emph{the RISC-V Week}, 2019.

\bibitem{10473983}
{Xingquan Li et al.}, ``{iEDA}: An open-source infrastructure of {EDA},'' in
  \emph{ASP-DAC)}, 2024.

\bibitem{soeken2018epfl}
M.~Soeken, H.~Riener, W.~Haaswijk, E.~Testa, B.~Schmitt, G.~Meuli, F.~Mozafari,
  S.-Y. Lee, A.~T. Calvino, D.~S. Marakkalage \emph{et~al.}, ``The {EPFL} logic
  synthesis libraries,'' \emph{arXiv preprint arXiv:1805.05121}, 2018.

\bibitem{neto2019lsoracle}
W.~L. Neto, M.~Austin, S.~Temple, L.~Amaru, X.~Tang, and P.-E. Gaillardon,
  ``{LSOracle}: A logic synthesis framework driven by artificial
  intelligence,'' in \emph{IEEE ICCAD}, 2019.

\bibitem{biagetti2023open}
G.~Biagetti, L.~Falaschetti, P.~Crippa, M.~Alessandrini, and C.~Turchetti,
  ``Open-source {HW/SW} co-simulation using {QEMU} and {GHDL} for {VHDL}-based
  {SoC} design,'' \emph{Electronics}, vol.~12, no.~18, 2023.

\bibitem{libreda}
{Thomas Kramer}, ``{LibrEDA},'' \url{https://libreda.org}, 2024.

\bibitem{khan2023ghazi}
Z.~R. Khan, W.~ul~Hasan, Z.~Rafique, A.~A. Ansari, and S.~R. Naqvi, ``{GHAZI}:
  An open-source {ASIC} implementation of {RISC-V} based {SoC},''
  \emph{Authorea Preprints}, 2023.

\bibitem{zhu2022greenrio}
Y.~Zhu, G.~Yin, X.~Wang, Q.~Yang, Z.~Luan, Y.~Zhang, M.~Wang, P.~Guo, X.~Wan,
  S.~Hu \emph{et~al.}, ``{GreenRio}: A modern {RISCV} microprocessor completely
  designed with a open-source {EDA} flow,'' in \emph{WOSET}, 2022.

\bibitem{khan2023ibtida}
M.~H. Khan, A.~A. Jalal, S.~Ahmed, A.~A. Ansari, and S.~R. Naqvi, ``{IBTIDA}:
  Fully open-source {ASIC} implementation of {Chisel}-generated system on a
  chip,'' \emph{Authorea Preprints}, 2023.

\bibitem{bender-git}
{PULP Platform contributors}, ``bender,''
  \url{https://github.com/pulp-platform/bender}, 2022.

\bibitem{morty-git}
------, ``morty,'' \url{https://github.com/pulp-platform/morty}, 2022.

\bibitem{svase-git}
------, ``{SV}ase,'' \url{https://github.com/pulp-platform/svase}, 2023.

\bibitem{sv2v-git}
{Zachary Snow}, ``sv2v,'' \url{https://github.com/zachjs/sv2v}, 2020.

\bibitem{lazy-synthesis}
W.~Y. et~al., ``Lazy man's logic synthesis,'' in \emph{IEEE/ACM ICCAD}, 2012.

\bibitem{zimmermann1998library}
R.~Zimmermann, ``{VHDL} library of arithmetic units,'' in \emph{Proc. First
  Int. Forum on Design Languages (FDL’98)}.\hskip 1em plus 0.5em minus
  0.4em\relax Citeseer, 1998.

\bibitem{wolfgang2002Booth}
W.~Paul and P.-M. Seidel, ``To {Booth} or not to {Booth},'' \emph{Integration},
  vol.~32, no.~1, 2002.

\bibitem{shazad2015Booth}
S.~Asif and Y.~Kong, ``Performance analysis of wallace and radix-4
  booth-wallace multipliers,'' in \emph{2015 Electronic System Level Synthesis
  Conference (ESLsyn)}, 2015, pp. 17--22.

\bibitem{barzen2023abc9}
B.~L. Barzen, A.~Reais-Parsi, E.~Hung, M.~Kang, A.~Mishchenko, J.~W. Greene,
  and J.~Wawrzynek, ``Narrowing the synthesis gap: Academic {FPGA} synthesis is
  catching up with the industry,'' in \emph{DATE}, 2023.

\end{thebibliography}
\end{document}